\documentclass{www2012-accepted}

\usepackage{url}
\usepackage{rotating}

\begin{document}

\clubpenalty=10000 
\widowpenalty = 10000

\title{CloudGenius: Decision Support for Web Server Cloud Migration}

%
%
%
%
%

\numberofauthors{2} 
%
\author{
%
%
\alignauthor
Michael Menzel\\
       \affaddr{Research Center for Information Technology}\\
       \affaddr{Karlsruhe Institute of Technology}\\
       \affaddr{Karlsruhe, Germany}\\
       \email{menzel@fzi.de}
\alignauthor
Rajiv Ranjan\\
       \affaddr{School of CSE, University of New South Wales}\\
       \affaddr{Sydney, Australia}\\
       \affaddr{Information Engineering Lab, CSIRO ICT Center}\\
       \affaddr{Canberra, Australia}\\
       \email{rajiv.ranjan@csiro.au}
}
\date{31 October 2011}
\maketitle
\begin{abstract}


Cloud computing is the latest computing paradigm that delivers hardware and software resources as virtualized services in which users are free from the burden of worrying about the low-level system administration details. Migrating Web applications to Cloud services and integrating Cloud services into existing computing infrastructures is non-trivial. It leads to new challenges that often require innovation of paradigms and practices at all levels: technical, cultural, legal, regulatory, and social. The key problem in mapping Web applications to virtualized Cloud services is selecting the best and compatible mix of software images (e.g., Web server image) and infrastructure services to ensure that Quality of Service (QoS) targets of an application are achieved. The fact that, when selecting Cloud services, engineers must consider heterogeneous sets of criteria and complex dependencies between infrastructure services and software images, which are impossible to resolve manually, is a critical issue. To overcome these challenges, we present a framework (called CloudGenius) which automates the decision-making process based on a model and factors specifically for Web server migration to the Cloud. CloudGenius leverages a well known multi-criteria decision making technique, called Analytic Hierarchy Process, to automate the selection process based on a model, factors, and QoS parameters related to an application. An example application demonstrates the applicability of the theoretical CloudGenius approach. Moreover, we present an implementation of CloudGenius that has been validated through experiments.  

\end{abstract}


\\

\section{Introduction}
The emergence of Cloud computing over the past five years is potentially one of the breakthrough advances in the history of computing. Cloud providers including Amazon Web Services (AWS), Salesforce.com, or Google App Engine give users the option to deploy their application over a network of infinite resource pool with practically no capital investment and with modest operating cost proportional to the actual use. By leveraging Cloud services to host Web applications organizations can benefit from advantages such as elasticity, pay-per-use, and abundance of resources. However, organisations tend to avoid or delay migrations of Web applications to the Cloud due to multiple hurdles. With Cloud computing being a disruptive technology an adoption brings along risks and obstacles \cite{armbrust2009above}. Risks can turn into effective problems or disadvantages for organizations that may decide to move Web applications to the Cloud. Considering this increases the complexity of the decision associated with migrating a Web application to the Cloud. Such a decision depends on many factors, from risks and costs to security issues and service level expectations \cite{SPE:SPE1110}. Another critical hurdle is the complexity of migrating a Web application to the Cloud on a technical level while incorporating economical aspects. A migration from an organization-owned data center to a Cloud infrastructure service implies more than few trivial steps. The following steps outline a migration of an organization's Web application to an equivalent on a Cloud infrastructure service. Steps of a migration to a Platform-as-a-Service (PaaS) offering would differ in several regards. 

First, an appropriate Cloud infrastructure service, or In\-fra\-struc\-ture-as-a-Service (IaaS) offering, is selected. This demands a well-thought decision to be made that considers all relevant factors, like e.g., price, Service Level Agreement (SLA) level, or support quality. The basis of a selection are data and measurements regarding each factor that describe the quality and make service options comparable. 
Secondly, the existing Web application and its platform, i.e., a Web server, are transferred from the local data center to the selected Cloud infrastructure service. Therefore, the Web application and server must be converted into a format expected by a Cloud infrastructure service. Typically, in this step the whole Web application is bundled as a virtual machine (VM) image that consists of a software stack from operating system and software platforms  to the software containing the business logic. Since it is often unachievable to convert an existing Web application and its server directly to a Cloud infrastructure service compatible VM image format, an adequate existing VM image offered by the Cloud provider is chosen and customized. Existing images vary in many ways, such as underlying operating system, software inside the software stack or software versions. Hence, selecting a functionally correct VM image becomes a complex  task. Besides, choosing a comprehensive VM image helps to minimize the effort of installing a software stack on a basic image. In the end, the resulting VM image should reflect the original Web server and at least replace it in a sufficing manner.
Next, a migration strategy needs to be defined and applied to make the transition from the local data center to the Cloud infrastructure service. A migration strategy defines the migration procedure in means of order and data transfer. In case data is affected with the Web application migration all data on the original machine must be transferred to the new system in the Cloud. Finally, all configurations and settings must be applied on the new Web server in the Cloud to finish creating an appropriate equivalent.
In sum, the process of migrating an IT system to a Cloud infrastructure service comprises five steps listed in presorted, modifiable order as following:
\begin{itemize}
\item Cloud infrastructure service selection
\item Cloud VM image selection
\item Cloud VM image customization
\item Migration strategy definition
\item Migration strategy application
\end{itemize}

The order reflects the fact that an image can be chosen for a certain Cloud infrastructure service only. Alternatively, selecting a Cloud VM image first restrains the number of eligible Cloud infrastructure services, typically to one.
In more complex settings multiple components and databases must be migrated in parallel, what requires to apply the steps described above component-wise. Additionally, interconnections and relations between the components must be considered.

In this paper we introduce the CloudGenius framework that lowers hurdles introduced by the complexity of the Cloud migration process. CloudGenius offers a detailed process and comprehensive decision support that reduces a Web engineer's effort of finding a proper infrastructure service and VM image when migrating a Web application to the Cloud. 

The paper is structured as follows. First, we reflect related work in Section \ref{related} and then present the CloudGenius framework in Section \ref{framework}. Further, we give an example application of CloudGenius in Section \ref{example} and present CumulusGenius, a prototypical implementation, in Section \ref{prototype}. In Section \ref{experiments} we present the results of experiments on CumulusGenius' time complexity before we discuss limitations and future work in Section \ref{discussion-future} and conclude in Section \ref{conclusion}.

\section{Related Work}\label{related}

There is preliminary work in the field of decision support for Cloud VM images and Cloud services.
Dastjerdi et al. \cite{dastjerdi2010effective} propose an approach that selects Cloud VM images and Cloud infrastructure services based on an ontology but lacks a service evaluation. Services that fulfill all requirements can not be differentiated regarding quality and suitability in this approach due to the missing evaluation. An ontology-based description to formulate requirements and attributes of a service is used since Dastjerdi et al. introduce a central system that manages service discovery and mediation.
Khajeh-Hosseini et al. \cite{khajeh2010cloud} \cite{khajeh2011decision} developed the Cloud Adoption Toolkit that offers a decision support for migrating a whole IT infrastructure of an enterprise. The focus of the decision support is on risk management and cost calculation based on workloads. However, the toolkit lacks considering factors that are not cost-related and supports a decision process which stays on an abstract decision level.
Regarding multi-component migration planning, Hajjat et al. \cite{hajjat2010cloudward} made some interesting effort with a model that supports decisions in hybrid Cloud setups. The model addresses the trade-off between direct cost savings, network delays, and internet communication costs and takes a set of specific constraints and requirements into account.
Ye et al. \cite{ye2011genetic} solve the mapping of service compositions to Cloud services with a genetic algorithm. The approach considers data and control flow of a composition and evaluates Cloud services by four Quality of Service (QoS) attributes. The approach overlooks the existence of Cloud VM images and focuses on service composition.

There is an amount of work on Cloud service selections that, however, lacks factoring in the actual Web server and VM image selection, and misses a decision support and a migration process. 
Chan and Chieu \cite{chan2010ranking} propose a single value decomposition technique-based approach that evaluates and ranks Cloud computing services by QoS attributes.
Li et al. \cite{li2010cloudcmp} present an approach to measure a provider's performance capabilities. The work introduces a set of interesting metrics to measure, but a method to choose a provider is missing. 

\section{CloudGenius Framework}\label{framework}

\begin{figure}[h]
\centering
\epsfig{file=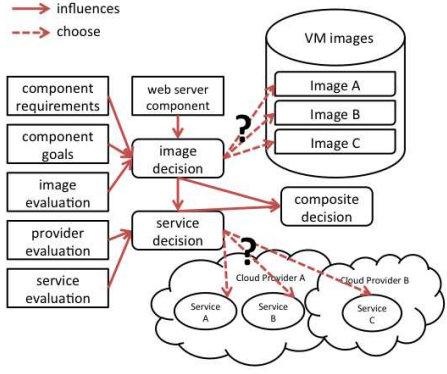, width=0.7\columnwidth}
\caption{Overview of the Selection Problem}\label{overview-selection}
\end{figure}

The complexity of a Web server Cloud migration can be mitigated by a decision support system that is capable of enhancing the quality of Cloud infrastructure service selections and Cloud VM image selections. Both selection tasks can be translated into decision-making problems between multiple alternatives. Each alternative, a Cloud infrastructure service or Cloud VM image, possesses different attributes which can be compared and evaluated with criteria. Finally, both selections must be joined since an image and a service build a single, combined solution. Usually, every Cloud VM image is only compatible with Cloud infrastructure services of few or a single provider restricting the number of viable combinations. Figure \ref{overview-selection} depicts the decisions to be made, as well as influential parameters and potential objects of choice. The problem depiction assumes that a user wants to choose a VM image before an infrastructure service. Solid lined arrows imply parameter inheritance and dashed lined arrows point at options to choose from.

\begin{figure}[h]
\centering
\epsfig{file=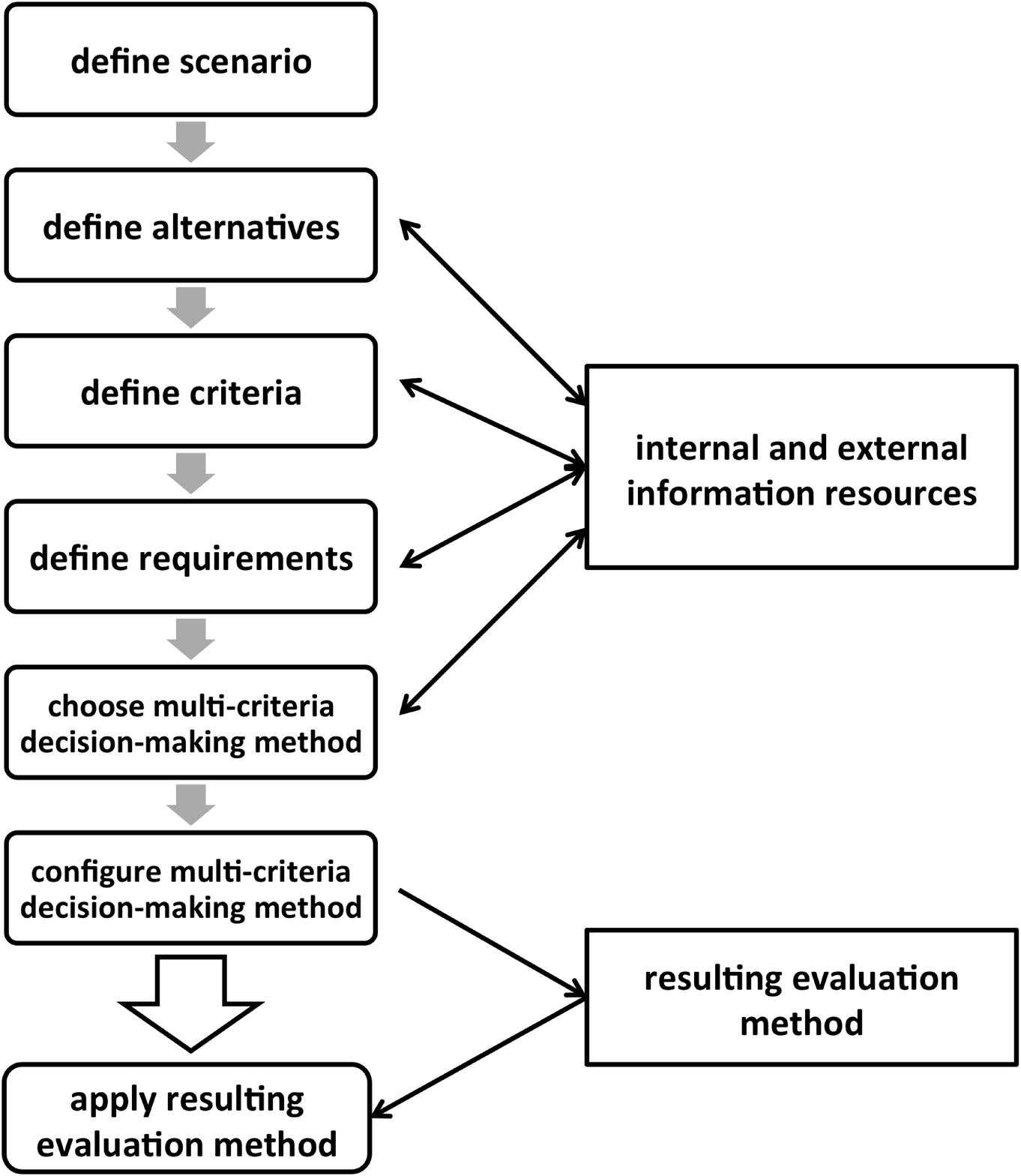, width=0.7\columnwidth}
\caption{Overview of the $(MC^2)^2$ Process}\label{mc22-process}
\end{figure}

With CloudGenius we propose an approach that translates both selection steps into multi-criteria decision-making problems to determine the most valuable combination of a Cloud VM image and a Cloud infrastructure service. The CloudGenius framework defines a Cloud migration process. Within the process CloudGenius offers a model and methods to determine the best combined choice of a Cloud VM image and a Cloud infrastructure service. The framework leverages an evaluation and decision-making framework, called $(MC^2)^2$, \cite{menzel2010} to support requirements and adopt a profound multi-criteria evaluation approach. The $(MC^2)^2$ framework provides a process depicted in Figure \ref{mc22-process} that allows to create an evaluation method that contains a requirements check and evaluates multiple alternatives with relative values on a (0-1) scale. Within $(MC^2)^2$ process CloudGenius proposes the Analytic Hierarchy Process (AHP) reducing decision modelling effort in comparison to the Analytic Network Process (ANP) suggested by $(MC^2)^2$. AHP allows to build complex weighted sum functions in a structured manner. Weighted sum functions allow compensation between criteria what can be influenced by weights derived from trade-offs stated by a user. 

CloudGenius allows users to define an abstract Web server and set requirements for a Web server implementation that condition what attributes are acceptable for a Cloud equivalent. Besides, a user has to choose relevant factors from a criteria list and define their priorities by setting weights for criteria in pair-wise comparisons. From the information given by the user the CloudGenius approach employs a model and the user preferences to apply the $(MC^2)^2$ framework and suggest a best Cloud image and Cloud service combination. 

The following subsections give details on CloudGenius' migration process, its formal model, and the selection and combination steps that lead to an adequate VM image and Cloud service from the abundance of offerings. Moreover, the final subsections address an alternative evaluation variant and the computational complexity of CloudGenius.

\subsection{A Cloud Migration Process}

\begin{sidewaysfigure}
\centering
\epsfig{file=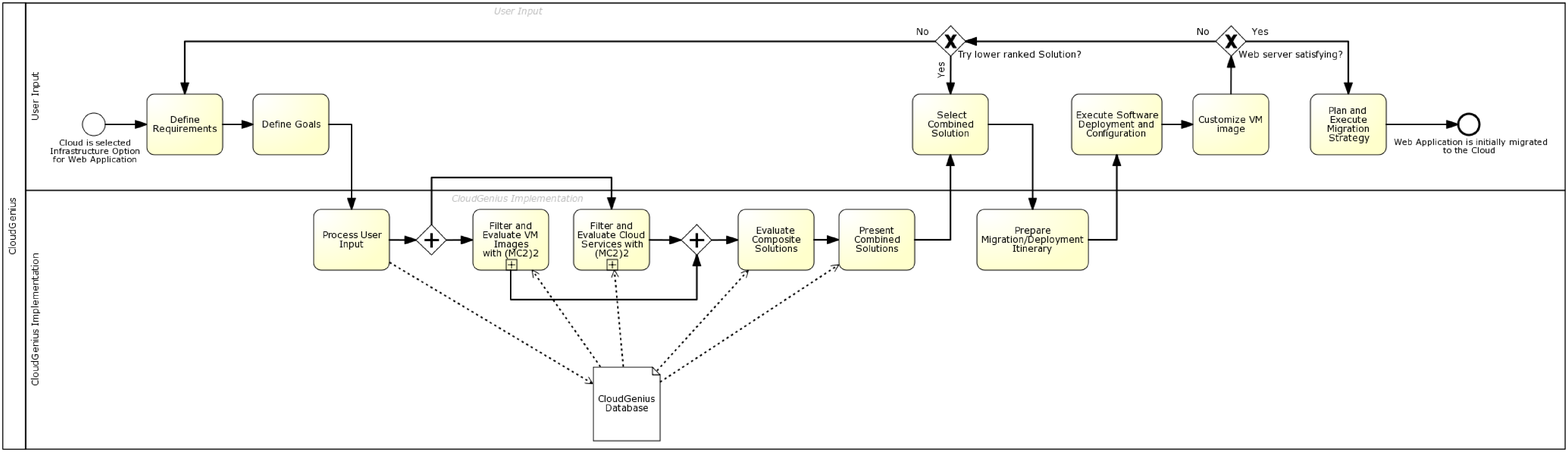, width=\textwidth}
\caption{Migration Process of the CloudGenius Framework}\label{cloudgenius-migration-process}
\end{sidewaysfigure}

Figure \ref{cloudgenius-migration-process} shows the migration process to be followed in the CloudGenius framework modelled in Business Process Model and Notation (BPMN) 2.0. The process illustration intends to define steps and highlight activities. The process is triggered by an event "Cloud is selected Infrastructure Option" occuring after an organization has identified the Cloud infrastructure for a possible or the best infrastructure option for a Web server. The subsequent steps show the information and preferences that need to be provided by the organization and Web engineer (see "User Input" lane) and what is processed and computed by CloudGenius resp. an implementation of the framework (see "CloudGenius implementation" lane). 
Two parallel steps that lie in the responsibility of CloudGenius are linking to the sub-process described by the $(MC^2)^2$ generic decision-making framework. Therefore, the implementation translates user input and the CloudGenius model stored in its database into a model that can be evaluated with $(MC^2)^2$. In the process Web engineers provide input and receive suggestions computed by CloudGenius. From there the migration is executed with a selected VM image and an infrastructure service target.

In the final steps of the process a user has the chance to enter a loop, try worse alternatives or reconsider with altered requirements and a new focus. Thereby the process allows users to incrementally improve a migration until a satisfying solution has been found. The migration process ends when the Web application has been successfully migrated to the Cloud.  

\subsection{Formal Model of CloudGenius}

In order to formalize the problem addressed by the CloudGenius framework, a formal mathematical model is introduced. The model defines the sets of objects involved in the problem and the relations between objects and their evaluation.
Table \ref{cloudgenius-math-model} summarizes the parameters of the model included in CloudGenius. 

\begin{table*}
\centering
\caption{CloudGenius Model}\label{cloudgenius-math-model}
\begin{tabular}{|l|p{4in}|} \hline
Parameter&Description\\ \hline
$w$&Web application to be migrated to the Cloud\\ 
$A = \{a_1,...,a_m\}$&set of $m$ Cloud VM images\\
$S = \{s_1,...,s_n\}$&set of $n$ Cloud infrastructure services\\
$P = \{p_1,...,p_o\}$&set of $o$ Cloud providers\\
$R_A = \{r_{A,1},...,r_{A,r_A}\}$&set of $r_A$ Cloud image requirements\\
$R_S = \{r_{S,1},...,r_{S,r_S}\}$&set of $r_S$ Cloud service requirements\\
$D = \{d_1,...,d_q\}$&set of $q$ image-service dependencies $(a_i, s_j)$\\
$\tau_h$&Cloud image $a_i$ or service $s_j$ as h-th $\tau \in A\cup{}S$\\
$\hat{A}_{\tau_h} = \{\alpha_{{\tau_h},1},...,\alpha_{{\tau_h},t}\}$&set of $t$ numerical attributes of h-th image or service\\
$\hat{B}_{\tau_h} = \{\beta_{{\tau_h},1},...,\beta_{{\tau_h},u}\}$&set of $u$ non-numerical attributes of h-th image or service\\
$\chi(\alpha)$&Value of numerical attribute $\alpha$ in CloudGenius database\\ 
$\chi(\beta)$&Value of non-numerical attribute $\beta$ in CloudGenius database\\ 
$v_{\tau_h}$&value of h-th $\tau$ calculated with $(MC^2)^2$\\
\hline\end{tabular}
\end{table*}


The model behind CloudGenius consists of $m$ images $a_i$, $n$ services $s_j$ and $o$ providers $p_k$. $A$ is the corresponding set of Web server images, $S$ the set of Cloud infrastructure services and $P$ the set of Cloud providers. $a_i$ and $s_j$ own numerical, measurable and non-numerical attributes defined in $\hat{A}_{a_i}$, $\hat{A}_{s_j}$, $\hat{B}_{a_i}$ and $\hat{B}_{s_j}$. $\chi$ represents a value connected with a numerical attribute $\alpha$ or non-numerical attribute $\beta$. Furthermore, the model introduces $r_A$ VM image related requirements and $r_S$ service related requirement, and a goal/criteria hierarchy including $g$ goals and $c$ leaf criteria. 

Based on the model CloudGenius determines the best combination ($a_i$, $s_j$) where $a_i$ and $s_j$ are the image and infrastructure service of provider $p_k$ that have the highest value of all combinations according to the user's preferences and with $a_i$ deployable on $s_j$. An evaluation method built with $(MC^2)^2$ can be employed to determine the best image $a_i \in A$. Therefore, $(MC^2)^2$'s function is interpreted as $f(a_i, \hat{A}_{a_i}, \hat{B}_{a_i})\mapsto{}v_{a_i}$ which returns a value $v_{a_i}$ for image $a_i$ and allows to find\\ $max \{v_{a_1}, ..., v_{a_m}\}$.
In parallel, the best service $s_j \in S$ has the value $max \{v_{s_1}, ..., v_{s_n}\}$ with $f(s_j, \hat{A}_{s_j}, \hat{B}_{s_j})\mapsto{}v_{s_j}$. In a final step, given the evaluation results of Cloud VM images and infrastructure services, a best combination is determined. Combinations of an image and a service must be feasible, meaning an image has to be deployable on the service. The feasibility of a combination is indicated by the set $D$ which holds all dependencies between images and services. Not all providers do support standard formats and, thus, set $D$ is required to define the compatibilities between images and services explicitly.

The applicability of the approach correlates with the quality of the data hold in the model. Therefore, the number of images and services must possibly reflect the Cloud market that is of interest for a Web server migration. Moreover, the currentness and quality of $\chi$ values of all numerical and non-numerical attributes have a high, direct influence on the quality of CloudGenius' results. Thus, measurements must be accurate and up-to-date implicitly demanding frequent maintenance of the database.

\subsection{Web Server Requirements \& Preferences}

In the first step of applying CloudGenius, a Web engineer has to define and describe the Web server by formulating requirements and their preferences. Requirements formulation comprises choosing a Web server attribute and specifying a minimum or maximum value for numerical values, and a set of allowed items for non-numerical values. Numerical requirements can be of Max or Min type, non-numerical requirements of Equals or OneOf type. Table \ref{requirement-types} gives an overview of the possible requirement definitions aligned with the $(MC^2)^2$ framework which uses conjunctive and disjunctive satisficing methods for numerical values. The table assumes $\chi$ to be the attribute value under consideration, $v_r$ to be the given numerical requirement value, $s$ a given non-numerical value and $S$ a given set of non-numerical values. Boolean expressions mathematically formulate applied requirements checks regarding an attribute value $\chi$ and the requirement constraint.
A list of available Web server attributes to set requirements upon can be drawn from merging all available attributes of Cloud VM images and Cloud infrastructure services listed in Table \ref{image-numerical-attributes}, \ref{image-nonnumerical-attributes}, \ref{service-numerical-attributes} and \ref{service-nonnumerical-attributes}. The tables includes numerical and non-numerical attributes that can be referred to in a requirement condition of a requirement with an according numerical or non-numerical type. 

\begin{table}
\centering
\caption{CloudGenius Requirement Types}\label{requirement-types}
\begin{tabular}{|c|c|l|} \hline
Value Type&Req. Type&Boolean Expression\\ \hline
Numerical&Max&$\chi(\alpha) < v_r$\\
Numerical&Min&$\chi(\alpha) > v_r$\\
Non-numerical&Equals&$\chi(\beta) = s$\\
Non-numerical&OneOf&$\chi(\beta) \in S$\\
\hline\end{tabular}
\end{table}

\begin{figure}[h]
\centering
\epsfig{file=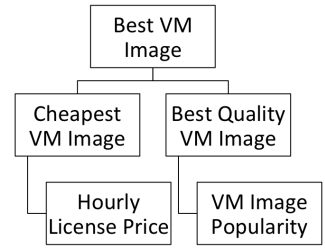, width=0.5\columnwidth}
\caption{CloudGenius VM Image Goal Hierarchy}\label{overview-image-goals}
\end{figure}

\begin{figure}[h]
\centering
\epsfig{file=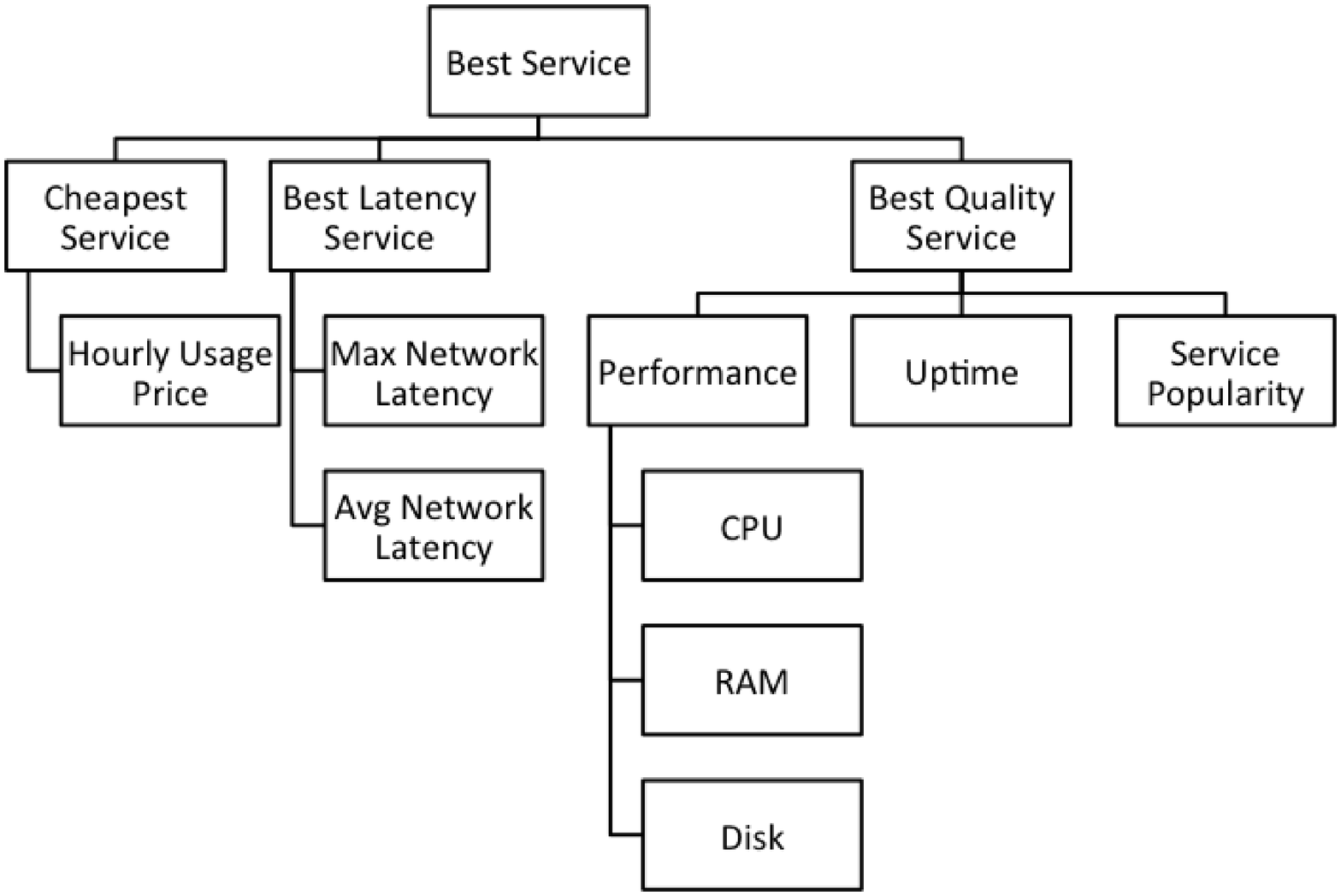, width=0.9\columnwidth}
\caption{CloudGenius Infrastructure Service Hierarchy}\label{overview-service-goals}
\end{figure}

Stating preferences is carried out by selecting and weighting goals and criteria. CloudGenius proposes a hierarchy of goals and criteria and asks a Web engineer to weight the items in pair-wise comparisons, in analogy to AHP. Figure \ref{overview-image-goals} and Figure \ref{overview-service-goals} depict the proposed hierarchy offered by CloudGenius. The proposed criteria for CloudVM images and Cloud infrastructure services are derived from Kalepu et al. \cite{kalepu2003verity}. For CloudVM images the goal hierarchy comprises two goals, cheapest and best quality VM images, with one criterion each, hourly price and image popularity.
The hierarchy for services consists of three main goals: cheapest, best latency and best quality service, that parent multiple criteria: hourly price, max. and average network latency, performance and uptime. The Performance goal has three sub-goals CPU, RAM and Disk performance. 
The criterion "popularity" corresponds to "reputation" and "uptime" to "availability" in Kalepu et al..

The selection of goals comprises few, important QoS attributes only, but CloudGenius is extensible regarding a wider selection of goals and more complex hierarchy. Since the $(MC^2)^2$ evaluation framework supports more complex goal hierarchies CloudGenius is not limited in this aspect. However, a higher complexity of the goal hierarchy increases the amount of input a user must provide. Our approach keeps the goal hierarchy complexity low but allows users to extend and customize the goal hierarchy to their needs. Possible additional goals are QoS properties proposed by Kalepu et al. \cite{kalepu2003verity} such as reliability, accessibility, security or support.

Finally, it is important to find a trade-off for the compensatory influence of Cloud VM image and Cloud infrastructure service on the combined value in a pair-wise comparison. The trade-off is incorporated in the calculation of an overall value of a combination of an image and a service.

\subsection{Cloud VM Image Selection}

A Cloud VM image is a pre-configured, self-contained, virt\-ual\-izat\-ion-enabled, and pre-built software resource unit that can be deployed at a Cloud infrastructure service. An example for a Web Server Cloud image is a pre-configured Tomcat Web server on Ubuntu Linux 10.4. Currenlty, there is no widely accepted standard (virtualization) format for a Cloud image, which means that an image is built for a specific virtualization technology and possibly will not run on Cloud services that are managed by non-overlapping virtualization technologies. For example, a VMware image cannot be deployed on the Amazon Web Services (AWS) Elastic Compute Cloud (EC2) service, as AWS only supports a proprietary virtualization format (also referred to as Amazon Machine Image (AMI) virtualization format).

Table \ref{image-numerical-attributes} gives an overview of the measurable, numerical attributes the CloudGenius framework proposes. The table lists their influence direction, metrics and the range of possible values. A list of non-numerical attributes is presented in Table \ref{image-nonnumerical-attributes}. The variability of an attribute represent the ability to change the value over time, e.g., the popularity of an image might shrink over time and, thus, is dynamic. Static variability means the value is defined once with the creation of the image, e.g., its operating system, while dynamic attribute values are more volatile. Non-numerical attributes are all of static variability, while all numerical attributes are dynamic. Nevertheless, there is no correlation between non-numerical attributes and static variability in general. In parallel, numerical attributes are not always dynamic per se. Attributes, additionally, have either a negative or positive influence on the value. For instance, the higher the license costs the less interesting becomes an image. In contrary, higher popularity is positive for an image. $(MC^2)^2$ respects negative and positive values in its evaluation.

\begin{table}[h]
\centering
\caption{Cloud Image Numerical Attributes}\label{image-numerical-attributes}
\begin{tabular}{|l|l|l|l|} \hline
Name&Influence&Metric&Value Range\\ \hline
Hourly License Price&Negative&\$/h&0-$\infty$ \$/h\\
Popularity&Positive&\%&0-100 \%\\
\hline\end{tabular}
\end{table}

\begin{table}[h]
\centering
\caption{Cloud image Non-numerical Attributes}\label{image-nonnumerical-attributes}
\begin{tabular}{|l|l|} \hline
Name&Example Values\\ \hline
Virtualization Format&Xen, VMWare, \ldots\\
Operating System (OS)&Linux, Windows, \ldots\\
OS Version&Ubuntu 10.4, \ldots\\
Implementation Language&Java, Perl, Ruby, \ldots\\
Supported Impl. Lang.s&Java, Perl, Ruby, \ldots\\
\hline\end{tabular}
\end{table}

The list of Cloud VM image attributes is open for attribute changes and supports users with the introduction of more or other attributes for images. The current attribute selection is derived from literature \cite{dastjerdi2010effective} and own observations, and provides a basic set of attributes that are essential to Cloud VM image selection. Nevertheless, the current Cloud VM image attribute list is not fixed and we hope to be able to build a list of important attributes from usage data of a future, publicly available prototype. Both measurable attributes have been derived from \cite{kalepu2003verity} since these two attributes have been found to be useful and applicable to VM images.

The VM image selection process requires the $(MC^2)^2$ evaluation framework to evaluate whether an image can deliver the requested functional and non-functional requirements, e.g., "supports Web applications implemented in Java" or "has a popularity of over 80\% according to user ratings". It is assumed that all VM images are Web servers what can be ensured by introducing an attribute that describes the VM image software stack and an requirement set automatically that checks if this attribute implies a Web server, i.e., an attribute "VM Image Feature" and a requirement "VM Image Feature equals 'Web server'". Every image that does not meet all formulated requirements is eliminated by $(MC^2)^2$ and not considered for composite solutions of an image and a service. Besides, the evaluation framework determines a value for an image and, therefore, involves all numerical image attributes. 

To create an evaluation method with $(MC^2)^2$ that provides the demanded capabilities a Web engineer's input and CloudGenius' formal model must be translated into input parameters for the $(MC^2)^2$'s evaluation method building process (see Figure \ref{mc22-process}). 
Therefore, in the building process the "Define scenario" step is skipped and in the subsequent step "Define alternatives" all images from a database which holds the set of images including all image attributes become the evaluation alternatives.
Further, all numerical attributes of images (see Table \ref{image-numerical-attributes}) are translated into a criteria hierarchy. The criteria hierarchy corresponds to the proposed goal hierarchy customized and weighted in pair-wise comparisons  in step 2 of the process. Any items within the hierarchy that have been deselected are removed from the hierarchy.
CloudGenius defines all requirements set on image attributes as the requirements to be considered by the evaluation method. The multi-criteria decision-making method chosen is every time AHP which needs to be configured with criteria weights.
The weighting of the criteria is inferred from goal preferences which have been gathered in pair-wise comparisons for all available attributes during the first steps of the CloudGenius process. 

Fed with images as alternatives, a criteria hierarchy, requirements and criteria weights a new evaluation method based on AHP is created with $(MC^2)^2$. The created evaluation method matches a function $f(a_i, \hat{A}_{a_i}, \hat{B}_{a_i})\mapsto{}v_{a_i}$ which is applied on all images $a_i$. By setting all $\chi(\alpha)$ as parameters in the function a value $v_{a_i}$ for every image $a_i$ can be determined. The function returns a value $v_{a_i} = 0$ whenever the attributes $\hat(B)_{a_i}$ of $a_i$ do not achieve all requirements. In detail, $(MC^2)^2$ creates the function described in Equation \ref{function-image-evaluation} where $w_j$ is the global criteria weight of j-th numerical attribute.
\begin{equation}\label{function-image-evaluation}
\left.
\begin{aligned}
f(a_i, \hat{A}_{a_i}, \hat{B}_{a_i}) &=
\begin{cases}
\sum\limits_{j=0}^{|\hat{A}_{a_i}|} w_j\chi(\alpha_{j,a_i}) & \forall r \in R_A: r=true\\
0 & else
\end{cases}\\
&\mapsto{} v_{a_i}
\end{aligned}
\right.
\end{equation}

In case, none of the images meets all requirements in a subsequent step all images that meet all but one requirement are considered. This procedure is repeated n times until a set of images is found that fulfil all but n requirements. The evaluation will still continue and return evaluation values. CloudGenius expects a created evaluation method  to support a requirements check that keeps images with any number of requirements in $R_A$ not met, e.g., for $n=1$ the evaluation method must keep all images with a single, arbitrary requirement in $R_A$ not met. This must not be achieved by removing one element from $R_A$ subsequently and apply the evaluation method with a smaller $R_A$ set. Instead the requirements check step within the evaluation method must allow n requirement checks to be false for any image. Only then the calculated results are represented by relative values between all image alternatives.

To determine the best image one needs to find the highest value alternative of all evaluation results $max \{v_{a_1},...,v_{a_m}\}$. A reasonable notion is to order all alternatives resp. images by value what can be achieved with an arbitrary sorting algorithm. Then, depending on the sorting direction, the best alternatives are the top for descending order or bottom ones for ascending order.

\subsection{Cloud Infrastructure Service Selection}

In parallel to the Cloud VM image selection, the selection of a Cloud infrastructure service proposes a range of numerical and non-numerical attributes and leverages the $(MC^2)^2$ framework to gain an evaluation method $g(s_j, \hat{A}_{s_j}, \hat{B}_{s_j})\mapsto{}v_{s_j}$ formulated in Equation \ref{function-service-evaluation} where $w_i$ is the global criteria weight of i-th numerical attribute.

\begin{equation}\label{function-service-evaluation}
\left.
\begin{aligned}
g(s_j, \hat{A}_{s_j}, \hat{B}_{s_j}) &=
\begin{cases}
\sum\limits_{i=0}^{|\hat{A}_{s_j}|} w_i\chi(\alpha_{i,s_j}) & \forall r \in R_S: r=true\\
0 & else
\end{cases}\\
&\mapsto{} v_{s_j}
\end{aligned}
\right.
\end{equation}

A list of numerical attributes for the Cloud service selection evaluation method is given in Table \ref{service-numerical-attributes}, a list of non-numerical attributes in Table \ref{service-nonnumerical-attributes}. The table of numerical attributes implies a dynamic variability for all attributes and presents their metric and value range. The table of non-numerical attributes implies a static variablity for all attributes and gives example values.

\begin{table}[h]
\centering
\caption{Cloud Service Numerical Attributes}\label{service-numerical-attributes}
\begin{tabular}{|l|l|l|l|} \hline
Name&Influence&Metric&Value Range\\ \hline
Hourly Price&Negative&\$/h&0-$\infty$ \$/h\\
CPU Perfomance&Positive&Flops&0-$\infty$ Flops\\
RAM Perfomance&Positive&Flops&0-$\infty$ Flops\\
Disk Perfomance&Positive&Flops&0-$\infty$ Flops\\
Max. Latency&Negative&ms&0-$\infty$ ms\\
Avg. Latency&Negative&ms&0-$\infty$ ms\\
Uptime&Positive&\%&0-100\%\\
Service Popularity&Positive&\%&0-100\%\\
\hline\end{tabular}
\end{table}

\begin{table}[h]
\centering
\caption{Cloud Service Non-numerical Attributes}\label{service-nonnumerical-attributes}
\begin{tabular}{|l|l|} \hline
Name&Example Values\\ \hline
Provider&Amazon, Rackspace, \ldots\\
Location Country&Germany, Australia, \ldots\\
\hline\end{tabular}
\end{table}

Cloud infrastructure services possess multiple numerical attributes that imply a caclulation or measurement. Hourly Usage Prices are typically provided from the provider or must be calculated with a provider's price model. More detailed cost schemes are provided by Klems et al. \cite{klems2009clouds} and Khajeh-Hosseini et al. \cite{khajeh2011decision}. To guarantee comparability of costs the attribute might need a corrected metric and needs to be renamed to "Total Costs" or "Monthly Costs" accordingly. Performance and latency attributes want to be measured with benchmarking tools as those are often not provided by Cloud providers \cite{lenk2011you}. The uptime of a service is a long-term experience that can be provided as guaranteed uptime by a provider or his SLA, or retrieved from user experiences. Also, service popularity can be gained from user experiences.

In parallel to the process of building the image selection evaluation method, the evaluation method building process of $(MC^2)^2$ must be followed (see Figure \ref{mc22-process}). With all parameters set (images as alternatives, a goal and criteria hierarchy, requirements and criteria weights) the $(MC^2)^2$ process can be completed and a new Cloud service evaluation method based on AHP is created. The Cloud service evaluations can be retrieved from applying the new evaluation method $g(s_j, \hat{A}_{s_j}, \hat{B}_{s_j})$. The highest ranked Cloud infrastructure alternative is $max \{v_{s_1}, ..., v_{s_n}\}$.

\subsection{Best Combination}

In the final step, after both, images and services, have been evaluated and became comparable, combined solutions are build. Every VM image $a_i$ can be combined with a service $s_j$ to create a possible solution pair $(a_i, s_j)$. The newly created solutions are not all actually feasible combinations and, hence, infeasible image-service compositions need to be filtered out. The CloudGenius model holds the set $D$ of all dependencies between images and services that indicate the compatibility of image $a_i$ with service $s_j$. Only solution pairs that have a corresponding dependency defined in the set $D$ are interpreted as feasible and have a value $v_{a_i,s_j} > 0$. Equation \ref{function-composite-evaluation} formulates the function $b$ that maps solution pairs to a value using an operator $\bullet$. 

\begin{equation}\label{function-composite-evaluation}
\begin{aligned}
v_{a_i,s_j} = b: (a_i,s_j) \mapsto{} f(a_i,\hat{A}_{a_i},\hat{B}_{a_i}) \bullet g(s_j,\hat{A}_{s_j},\hat{B}_{s_j})\\
\forall i, j:  (a_i,s_j) \in D
\end{aligned}
\end{equation}

CloudGenius proposes $\bullet$ to be the operator $+$ and, hence, allows for compensation between the value of a VM image and a service within total value of a combination. For instance, a low quality service and high quality image might be a less good combination than a combination of medium quality image and medium quality service. Furthermore, user defined weights $w_a$ and $w_s$ that sum up to 1 determine the influence of $f$ and $g$ on the total value. Equation \ref{function-composite-evaluation-weights} formulates the resulting evaluation function suggested by CloudGenius. For more complex combination value computations a function $h(f(\cdot),g(\cdot))$ may return the overall value instead of an operator $\bullet$ only. A wider choice of operators or functions is subject to future enhancements and implementations, e.g., $*$ or $\times$ operator helps finding VM image infrastructure service combinations with most balanced values.

\begin{equation}\label{function-composite-evaluation-weights}
\begin{aligned}
b(a_i,s_j) &= 
\begin{cases}
w_a * f(\cdot) + w_s * g(\cdot) & (a_i,s_j) \in D\\
0 & else\\
\end{cases}\\
&\mapsto{} v_{a_i,s_j}
\end{aligned}
\end{equation}

The best overall solution pair to choose has the value:
\begin{equation*}\max b(a_i,s_j) = \max \{v_{a_1,s_1},...,v_{a_1,s_1}\}\end{equation*}
The ranking of only the VM image or the service of a solution within the list of evaluated images or services is an interesting indicator reflecting the compensation effect. With the selection of a combined solution the process continues. The image can be deployed on the service and further customized. The process ends with the execution of a migration strategy that results in a Web application available on Cloud infrastructure services.

\subsection{Integrated Evaluation Approach}

\begin{figure}[h]
\centering
\epsfig{file=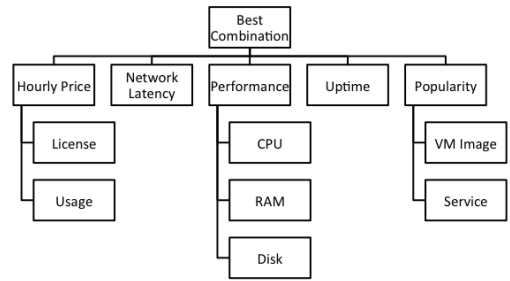, width=0.8\columnwidth}
\caption{CloudGenius Integrated Goal Hierarchy}\label{overview-goals-integrated}
\end{figure}

Alternatively, a best combination can be determined in a single $(MC^2)^2$ process using the goal hierarchy and according weights from user input depicted in Figure \ref{overview-goals-integrated}. In this variant the compensatory relation between VM image and service is less transparent and can no longer be influenced directly, but is by weighting the goal hierarchy. Hence, we suggest not to use the integrated approach and leave it to an implementation to provide more evaluation options to a user.

\subsection{Complexity of CloudGenius Approach}

The whole problem addressed by CloudGenius seems rather complex and involves a number of calculations. Hence, analyzing the actual computational complexity of the approach becomes of interest to ensure its applicability. To our best knowledge we define $O$ of CloudGenius as following:
\begin{align*}
O(
&\underbrace{m*|\hat{B_a}|+n*|\hat{B_s}|}_{\text{requirements check}} + \underbrace{m*|\hat{A_a}|+n*|\hat{A_s}|}_{\text{evaluation}} +\\
&\underbrace{m*n*|D|}_{\text{feasibility check}} + \underbrace{m*n}_{\text{combined value}})
\end{align*}

The computational complexity is proportional to the the number of VM images $m$, services $n$. Computations for $m$ images and $n$ services comprise requirements checks in $\hat{B_a}$ and $\hat{B_s}$, evaluations regarding criteria $\hat{A_a}$ and $\hat{A_s}$, feasibility check with set $D$, and combining and evaluating image-service pairs. Evaluations implicate additional computation steps for AHP comprising normalization of matrices and derivation of global weights which are not included for simplicity. Based on the given $O$ we expect CloudGenius' time complexity to be quadratic in proportion to $m$ and $n$.

\section{Example Application}\label{example}

The following example shows CloudGenius' support for selection and its process loop assuming a rich implementation and database.

An e-business relies on own IT infrastructures and suffers from high costs maintaining old systems. Thus, the e-business decides to use a Cloud infrastructure for its Web shop and applies the CloudGenius framework for migration. A Web engineer of the e-business has developed a scalable Web application providing a PHP environment that keeps data in the local, privately owned data center. Data migration, hence, is out of focus. Now, she defines requirements, and selects and weights goals and criteria. Her only requirement demands PHP to be supported as implementation language.
Next, she wants all proposed criteria to be considered and weights all hierarchies with an emphasis on latency and costs. With the given input the rich CloudGenius implementation starts the selection algorithm and suggests a combination of a Windows-based  VM image with a XAMPP\footnote{An Apache httpd, Perl, PHP, and MySQL software stack. See http://www.apachefriends.org/en/xampp.html, visited 2011-10-19.} stack pre-installed and an infrastructure service from Amazon. However, after deploying the VM image, and deploying and customizing the Web shop application on the software stack several problems occur that can be traced back to Windows-incompatibilities of the Web application.
Now, a new CloudGenius cycle begins and the Web engineer intends to pick a lower ranked non-Windows-based solution. Since the list is showing to many options she decides to step back to the requirements definition and adds a requirement which defines the Operating System to be "Linux". In the subsequent steps the CloudGenius implementation refilters all alternative VM images and offers an updated best combination list from which the top best alternative is chosen. From here the customization and installation of the Web application works properly. Finally, a migration strategy is planned that replaces the old, local Web application version. Fortunately, due to the absence of a data migration the migration strategy only comprises adding the new Web application's URL to an existing load balancer and remove old URLs over time.


\section{CumulusGenius Prototype}\label{prototype}

The complexity of the selection problem leads to the need of a software implementation that is fundamental to make CloudGenius applicable with acceptable effort. Therefore, we promote the CumulusGenius software tool \cite{cumulusgenius2011} that fully implements the model of VM images, services and providers, as well as requirements and criteria. For computing evaluations the Aotearoa software tool's API \cite{SPE:SPE1110}\cite{aotearoa2011} is used. Automated test deployments of images to infrastructure service mappings are triggered with the jCloud library \cite{jclouds2011}.
The current version of the CumulusGenius software tool implementation supports Web servers migration to Amazon Web Services (AWS) including all criteria presented earlier. The tool can be used programmatically as a java library and allows a user to create a data model with his requirements, criteria, and weights as well as attribute values of AMIs and services. The library expects the user to leverage enumeration classes that restrict definitions to supported requirements and criteria. The set of AMIs must be defined programmatically.

\section{Experiments}\label{experiments}

The theoretical computational complexity has already been determined earlier and has shown how multiple parameters of the formal model influence the complexity. The following experiment measure the actual time complexity of CumulusGenius and make a comparison to the expected $O$. 

\begin{figure}[h]
\centering
\epsfig{file=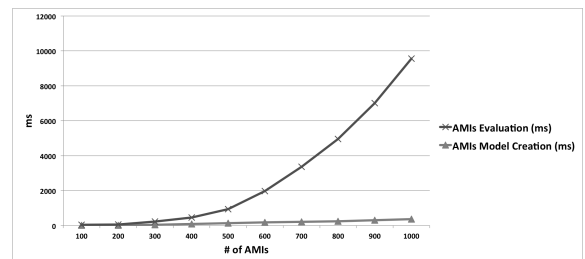, width=0.9\columnwidth}
\caption{Time Effort for AMI Selection}\label{experiments-ami-selection}
\end{figure}

\begin{figure}[h]
\centering
\epsfig{file=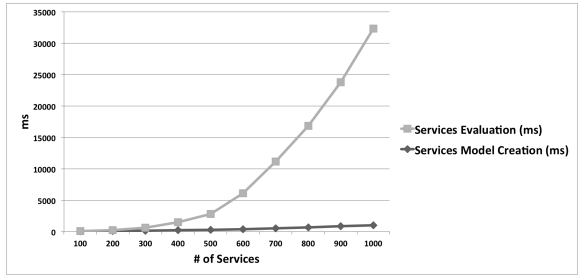, width=0.9\columnwidth}
\caption{Time Effort for Service Selection}\label{experiments-service-selection}
\end{figure}

The experiments were run in a 64Bit Java Virtual Machine version 1.6 on a MacBook Pro 5,5 with an Intel Core 2 Duo 2,5 Ghz CPU and 4GB of RAM. The experiment setup did not consider any requirements and skipped any filtering of infeasible combinations from a set $D$ what increases the total number of evaluations but eliminates filtering efforts. In every experiment the number of VM images and infrastructure service has be increased by 100 starting from 100 up to 1000 objects. Images and services were generated and random, plausible attribute values within value ranges were assigned. Figures \ref{experiments-ami-selection} and \ref{experiments-service-selection} depict the average time complexities for generating the model and evaluating $m$ images resp. $n$ services repeated in 20 runs. The experiments have shown that service evaluation is more complex than image evaluation originating from the higher number of criteria in service evaluations. 

\begin{figure}[h]
\centering
\epsfig{file=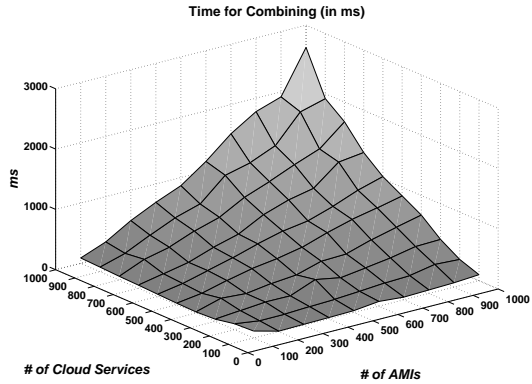, width=0.9\columnwidth}
\caption{Time Effort for Combinations}\label{experiments-combinations}
\end{figure}

Figure \ref{experiments-combinations} depicts the measured average time complexity of generating service-image combinations and the respective value from 20 runs for 100 to 1000 images and services. For combined values, both, images and services, have similar influence, but the computation effort grows non-linear. 

\begin{figure}[h]
\centering
\epsfig{file=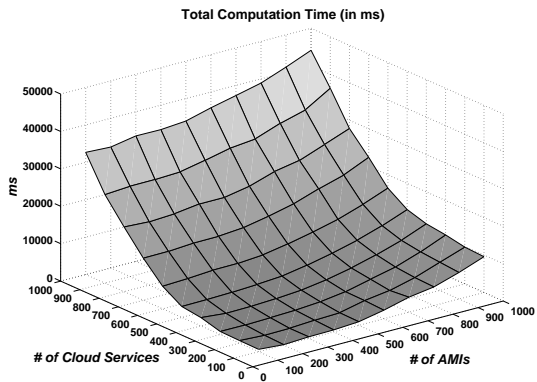, width=0.9\columnwidth}
\caption{Total Time Effort}\label{experiments-totals}
\end{figure}

The aggregated, total time complexity of a best combination computation depicted in Figure \ref{experiments-totals} reflects the non-linear behavior and the strong influence of a growing service number. In 20 runs 100 to 1000 images and services have been evaluated separately, grouped into combined solutions and an overall value.

The results confirm a quadratic time complexity and show that for a growing number of parameters, i.e., criteria, images, services, a search for a best solution is exhaustive. For large sets of parameters heuristics are an alternative approach. A first notion is a genetic algorithm employing function $b$ (see Formula \ref{function-composite-evaluation-weights}) as fitness function.

\section{Discussion \& Future Work}\label{discussion-future}


CloudGenius focuses on a migration process for single-tier Web applications and is limited in its applicability for other IT systems. Characteristics in means of criteria and process steps among others are inherent to an approach for Web server migration. A multi-tier decision support would have to consider relations between components and a persistence tier, all having distinct characteristics. Since the effort to evaluate qualitative criteria in pair-wise comparisons for all VM images and infrastructure services would be immense, besides non-numerical criteria the set of criteria is restricted to quantitative numerical criteria. Moreover, CloudGenius assumes a Web engineer expects a full featured VM image instead of a basic image with customization effort, such as a basic Linux OS image. 

However, the presented generic framework leaves space for a range of enhancements and deeper evaluations, and, yet, provides an applicable approach. To our knowledge no existing approach has addressed the problem of selecting a VM image and infrastructure service in a migration process.
Major issues are criteria catalogs, the quality and currentness of measured values, and a user-friendly implementation with a comprehensive database. To gain a comprehensive database existing databases such as thecloudmarket.com \cite{cloudmarket2011} can be integrated as well as  existing benchmarking services (e.g., CloudHarmony \cite{cloudharmony2011}) that together with automated benchmarking \cite{haak2011autonomic} mitigate staleness of data. Also, an integration of other tools, e.g., for different cost calculations, would add value to the CumulusGenius prototype. Moreover, the current complexity of requirement checks is very basic and might be extended with other approaches such as feature models \cite{wittern2011use}.
Regarding support for different VM image preparation levels future work can comprise a prediction of expected customization effort to compare full feature images with low featured images.
CloudGenius' limitation to quantitative criteria might be overcome with group ratings or evaluations of qualitative criteria on a rating portal in future prototype versions.
Future work also includes extending the support to more application types and to multiple, inter-related components. Besides, support for a persistence layer and system trade-offs such as the CAP theorem or security vs. latency as well as a wider process loop that begins with an initial Cloud decision are currently studied. Furthermore, the process itself might be enhanced with quality guarantees, e.g., guaranteed idempotency.

\section{Conclusion}\label{conclusion}

Decision problems occuring during the migration of Web applications to the Cloud are non-trivial. A decision support can decrease effort and remove hurdles.
In this paper, we introduced the generic CloudGenius framework that provides a migration process and decision support. With CloudGenius Web engineers are able to migrate Web applications to the Cloud along a cyclic process that suggests Cloud VM images and Cloud infrastructure services according to an engineers' requirements and goals. We further described limitations of the framework and gave an example to provide practical insights and show its applicability. In experiments a prototypic implementation of the framework has been validated and the time complexity of the underlying algorithm was analyzed.


\section{Acknowledgments}
Initial research work on Cloud VM image and infrastructure service data models was done when Dr. Rajiv Ranjan was employed at University of New South Wales on strategic eResearch grant scheme. 

%
\bibliographystyle{abbrv}
\bibliography{../../references}  

\begin{thebibliography}{10}

\bibitem{aotearoa2011}
{Aotearoa Prototype}.
\newblock \url{http://code.google.com/p/aotearoadecisions/}, accessed
  2011-10-19.

\bibitem{armbrust2009above}
M.~Armbrust, A.~Fox, R.~Griffith, A.~Joseph, R.~Katz, A.~Konwinski, G.~Lee,
  D.~Patterson, A.~Rabkin, I.~Stoica, et~al.
\newblock Above the clouds: A berkeley view of cloud computing.
\newblock {\em EECS Department, University of California, Berkeley, Tech. Rep.
  UCB/EECS-2009-28}, 2009.

\bibitem{chan2010ranking}
H.~Chan and T.~Chieu.
\newblock {Ranking and mapping of applications to cloud computing services by
  SVD}.
\newblock In {\em {Network Operations and Management Symposium Workshops (NOMS
  Wksps), 2010 IEEE/IFIP}}, pages 362--369. IEEE, 2010.

\bibitem{cloudharmony2011}
{CloudHarmony}.
\newblock \url{http://cloudharmony.com}, accessed 2011-10-19.

\bibitem{cumulusgenius2011}
{CumulusGenius Prototype}.
\newblock \url{http://code.google.com/p/cumulusgenius/}, accessed 2011-11-06.

\bibitem{dastjerdi2010effective}
A.~Dastjerdi, S.~Tabatabaei, and R.~Buyya.
\newblock An effective architecture for automated appliance management system
  applying ontology-based cloud discovery.
\newblock In {\em Proceedings of the 2010 10th IEEE/ACM International
  Conference on Cluster, Cloud and Grid Computing}, pages 104--112. IEEE
  Computer Society, 2010.

\bibitem{haak2011autonomic}
S.~Haak and M.~Menzel.
\newblock Autonomic benchmarking for cloud infrastructures: an economic
  optimization model.
\newblock In {\em Proceedings of the 1st ACM/IEEE workshop on Autonomic
  computing in economics}, pages 27--32. ACM, 2011.

\bibitem{hajjat2010cloudward}
M.~Hajjat, X.~Sun, Y.~Sung, D.~Maltz, S.~Rao, K.~Sripanidkulchai, and
  M.~Tawarmalani.
\newblock {Cloudward bound: Planning for beneficial Migration of Enterprise
  Applications to the Cloud}.
\newblock {\em {ACM SIGCOMM Computer Communication Review}}, 40(4):243--254,
  2010.

\bibitem{jclouds2011}
{jClouds Multi-Cloud Library}.
\newblock \url{http://code.google.com/p/jclouds/}, visited 2011-10-19.

\bibitem{kalepu2003verity}
S.~Kalepu, S.~Krishnaswamy, and S.~Loke.
\newblock Verity: a qos metric for selecting web services and providers.
\newblock In {\em Web Information Systems Engineering Workshops, 2003.
  Proceedings. Fourth International Conference on}, pages 131--139. IEEE, 2003.

\bibitem{khajeh2010cloud}
A.~Khajeh-Hosseini, D.~Greenwood, J.~Smith, and I.~Sommerville.
\newblock {The Cloud Adoption Toolkit: Supporting Cloud Adoption Decisions in
  the Enterprise}.
\newblock {\em Software: Practice and Experience}, 2010.

\bibitem{khajeh2011decision}
A.~Khajeh-Hosseini, I.~Sommerville, J.~Bogaerts, and P.~Teregowda.
\newblock Decision support tools for cloud migration in the enterprise.
\newblock {\em Arxiv preprint arXiv:1105.0149}, 2011.

\bibitem{klems2009clouds}
M.~Klems, J.~Nimis, and S.~Tai.
\newblock Do clouds compute? a framework for estimating the value of cloud
  computing.
\newblock {\em Designing E-Business Systems. Markets, Services, and Networks},
  pages 110--123, 2009.

\bibitem{lenk2011you}
A.~Lenk, M.~Menzel, J.~Lipsky, S.~Tai, and P.~Offermann.
\newblock What are you paying for? performance benchmarking for
  infrastructure-as-a-service offerings.
\newblock In {\em Cloud Computing (CLOUD), 2011 IEEE International Conference
  on}, pages 484--491. IEEE, 2011.

\bibitem{li2010cloudcmp}
A.~Li, X.~Yang, S.~Kandula, and M.~Zhang.
\newblock Cloudcmp: comparing public cloud providers.
\newblock In {\em Proceedings of the 10th annual conference on Internet
  measurement}, pages 1--14. ACM, 2010.

\bibitem{menzel2010}
M.~Menzel, M.~Sch{\"o}nherr, J.~Nimis, and S.~Tai.
\newblock {$(MC^2)^2$: A Generic Decision-Making Framework and its Application
  to Cloud Computing}.
\newblock In {\em Proceedings of the International Conference on Cloud
  Computing and Virtualization (CCV 2010)}, Singapore, Mai 2010. GSTF.

\bibitem{SPE:SPE1110}
M.~Menzel, M.~Sch{\"o}nherr, and S.~Tai.
\newblock {$(MC^2)^2$: Criteria, Requirements and a Software Prototype for
  Cloud Infrastructure Decisions}.
\newblock {\em Software: Practice and Experience}, 2011.

\bibitem{cloudmarket2011}
{The Cloud Market}.
\newblock \url{http://cloudmarket.com}, accessed 2011-10-19.

\bibitem{wittern2011use}
E.~Wittern and C.~Zirpins.
\newblock On the use of feature models for service design: the case of value
  representation.
\newblock In {\em Towards a Service-Based Internet. ServiceWave 2010
  Workshops}, pages 110--118. Springer, 2011.

\bibitem{ye2011genetic}
Z.~Ye, X.~Zhou, and A.~Bouguettaya.
\newblock Genetic algorithm based qos-aware service compositions in cloud
  computing.
\newblock In {\em Database Systems for Advanced Applications}, pages 321--334.
  Springer, 2011.

\end{thebibliography}
%
%

\balancecolumns
\end{document}